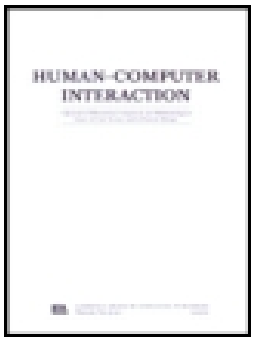

# Human–Computer Interaction

# Revisiting embodiment for brain–computer interfaces

**Barış Serim, Michiel Spapé & Giulio Jacucci**

Published online: 13 Feb 2023.

Submit your article to this journal 🗗

Article views: 49

View related articles 🗗

View Crossmark data 🗗

Taylor & Francis
Taylor & Francis Group

# Revisiting embodiment for brain–computer interfaces

Barış Serim [a], Michiel Spapé [b], and Giulio Jacucci [a]

[a]Department of Computer Science, University of Helsinki, Finland; [b]Department of Psychology and Logopedics, University of Helsinki, Finland

**ABSTRACT**

Researchers increasingly explore deploying brain–computer interfaces (BCIs) for able-bodied users, with the motivation of accessing mental states more directly than allowed by existing body-mediated interaction. This motivation seems to contradict the long-standing HCI emphasis on embodiment, namely the general claim that the body is crucial for cognition. This paper addresses this apparent contradiction through a review of insights from embodied cognition and interaction. We first critically examine the recent interest in BCIs and identify the extent cognition in the brain is integrated with the wider body as a central concern for research. We then define the implications of an integrated view of cognition for interface design and evaluation. A counterintuitive conclusion we draw is that embodiment per se should not imply a preference for body-grounded interaction over BCIs. It can instead guide research by 1) providing body-grounded explanations for BCI performance, 2) proposing evaluation considerations that are neglected in modular views of cognition, and 3) through the direct transfer of its design insights to BCIs. We finally reflect on HCI's understanding of embodiment and identify the neural dimension of embodiment as hitherto overlooked.

## 1. Introduction

Embodiment, the general claim that cognition is deeply dependent upon the physical structure of the body, influenced how HCI researchers theorize about and design user interfaces. For interface design, the concept of embodiment brought about an increased sensitivity toward the physical form of interfaces and the particular body movements performed by users. Researchers utilized the concept to underpin their research on physical gestures, tangible and ubiquitous interfaces and called for involving a more diverse set of physical forms and body gestures than required in traditional GUI interaction (Dourish, 2004; Hornecker & Buur, 2006; Ishii & Ullmer, 1997; Klemmer et al., 2006). Distinctions between physical and virtual, and between tangible and desktop computing have so far remained at the foreground of HCI discussions on embodiment.

At the same time, HCI is currently witnessing the emergence of brain–computer interfaces (BCIs). BCIs enable users to interact with computers without motor execution, namely without having to carry out body movements such as moving hands or fingers. Users with limited or no muscle control, such as those with impaired neural pathways due to brain injury, brainstem stroke, multiple sclerosis or any other condition, are thus an obvious target for BCI applications. Yet researchers increasingly develop BCIs for use by the general population. There are multiple motivations behind this recent interest such as the promise of brain signals as high-bandwidth communication channels (Schalk, 2008) or as accurate representations of mental content (Spapé et al., 2015).

The potential use of BCIs by the general population implies a future in which body movements could at least be partially replaced by brain signals, which represents a transformation that is at least as significant as the ongoing shift into mobile, ubiquitous and full-body interaction. Unlike these domains, however, BCIs received relatively little attention within the HCI discussions of embodiment. The elimination of body movements, at first sight, seems to go against the research program of embodied interaction, but some qualities that are attributed to ubiquitous or tangible interfaces, such as their foundation on familiar, everyday human skills and abilities (Abowd & Mynatt, 2000; Dourish, 2004; Jacob et al., 2008), also motivate the research on BCIs (e.g., Fairclough, 2011; Schalk, 2008). We believe that the time is due to revisit the concept of embodiment and its implications.

This paper thus examines the implications of embodiment for BCIs and uses the questions raised by BCIs to reflect on HCI's understanding of embodiment. We target the following audiences:

The first audience we target is researchers that study interaction modalities. Despite the recent interest, HCI has so far lacked an in-depth discussion on how BCIs would fit into the larger landscape of human–computer interaction. As researchers working on BCIs and other input modalities, we are preoccupied with questions such as: How will BCIs compare to existing input methods? What will be the relevance of the body and body movements for human–computer interaction in an era of viable BCIs? Should BCIs replace or complement body movements? These are essentially empirical questions for which we do not claim to have definitive answers. We instead provide methodological guidance to inform future work.

Our second target audience is HCI researchers working on BCIs. The expansion of research scope into the general population requires paying attention to how cognition in the brain is facilitated by and reflected in the wider body—an area of study that falls within the scope of embodied cognition and has so far received little attention within BCI and brain imaging research (Gramann et al., 2014). We address this knowledge gap by presenting evidence from embodied cognition and identifying different ways embodiment can guide research. The approach in this paper can thus be described as top-down. Instead of departing from the existing body of BCI work, we depart from the methodological and empirical insights of embodiment to identify research implications. We refer to BCI examples along the way for illustration.

Finally, the paper is intended for researchers with a general interest in the questions of embodiment and embodied interaction. Embodied interaction literature in HCI has drawn from diverse intellectual traditions and contains tensions concerning the design implications of embodiment and attitudes toward mental phenomena. The emergence of BCIs provides an opportunity for HCI to address these tensions and sharpen its own understanding of the concept.

Our contributions can be summarized as follows.

- We propose novel research directions for BCIs based on insights from embodiment and embodied cognition. We describe how embodied views of cognition lead to different research questions and desiderata than those elicited by modular views of cognition. We present these research directions with example study designs and identify methodological implications.
- We use the conceptual challenges posed by BCIs to further the understanding of embodiment in HCI and derive research implications for HCI at large. We first evaluate different claims within embodiment regarding their relevance for BCI research. We then reflect on the conceptual tensions within HCI and identify various ways an expanded understanding of embodiment can inform future research.

## 1.1. *Overview and organization*

Section 2 introduces the concept of "research vision" in HCI and identifies providing more direct access to mental states than enabled by existing input devices as a motivation for BCIs. It then formulates the concept of directness in terms of information transmission criteria to assess the promise of BCIs.

Section 3 investigates whether cognitivism provides a more explicit basis for the motivation of more direct access to mental states. We compare the commitments of cognitivism and embodiment regarding 1) the status of mental representations, 2) the relationship between meaning and interaction, 3) the neural basis of cognition, and 4) the extent cognition is integrated. We identify the fourth discussion point, the extent cognition is integrated, as the most relevant for the motivation of providing more direct access to mental states.

Section 4 proposes novel research directions for BCIs based on an integrated view of cognition. We present evidence related to embodied cognition and illustrate potential research directions by providing example study designs. We list a number of methodological implications that stem from an integrated view of cognition.

Section 5 reflects on HCI's understanding of embodiment. Taking cues from parallel discussions in embodied cognition, we argue that BCIs require defining embodiment in a way that does not limit its scope to body movements. We identify the wider implications of an expanded view of embodiment for HCI research.

Section 6 summarizes the main takeaways of the paper.

## 2. BCIs and a new research vision

Put simply, a brain–computer interface is a "communication system that does not depend on the brain's normal output pathways of peripheral nerves and muscles" (Wolpaw et al., 2000, p. 165). The current technical landscape of BCIs can be described as a plethora of different sensing methods (He et al., 2013) that are used in a diverse set of interactive applications ranging from real-time control to longitudinal user monitoring. Mental imagery-based BCI applications require users to control an interface through specific mental tasks. BCIs in this group, for instance, can ask participants to imagine left and right movements and use the time/frequency transformation of the EEG to predict the imagined movements, resulting in a joystick-like interaction element (Williamson et al., 2009). BCIs based on event-related potentials (ERPs) or steady-state visual-evoked potentials (SSVEPs), on the other hand, work by gathering brain activity while a user is subjected to visual or other sensory stimuli. For example, in P300 BCI spellers (Farwell & Donchin, 1988) users are tasked to focus on letter flashes; the enhanced evoked activity is then used to predict when a letter has been focused on, resulting in the letter to be spelled out by the system. Another BCI group targets using brain signals that occur as part of another activity. An example is the use of specific frequencies of the EEG spectrum to infer the "mental workload" of users and make various system adaptations in the background (Afergan et al., 2014; Yuksel et al., 2016).

Current BCI applications are generally targeted at enabling communication for people who have physical disabilities that restrict other types of input (e.g., Townsend et al., 2010; Williamson et al., 2009). In these use cases, BCIs are easily justified as they can be the only means for input. At the same time, the potential utility of BCIs as an everyday input for the wider population is increasingly being investigated in academic (e.g., Allison et al., 2007; Blankertz et al., 2010, 2016; Erp van et al., 2012), military (Darpa, 2018) and commercial contexts (Levy, 2017). In contrast to disability use cases, the deployment of BCIs to the wider population implies replacing manual input devices that currently dominate HCI, a shift that implies certain advantages for BCIs. These advantages are not yet demonstrated; the performance of the state-of-the-art BCIs falls short of established input methods for elementary tasks such as text entry or pointing (Pandarinath et al., 2017). The current research interest instead seems to be driven by expectations of future improvements in performance, which consequently points to a *theoretical potential* that exceeds what is possible with state-of-the-art techniques. It is then necessary to identify how this expectation is formulated and what constitutes the *research vision* for BCIs.

## 2.1.  *A research vision of directly accessing mental states*

Research visions can be defined as principles that guide a research enterprise beyond the limitations of current-day technological capabilities (Ishii et al., 2012). As such, research visions themselves are drivers of technological innovation and help organize individual research contributions as part of a coherent program. Among well-known HCI research visions are Bush's concept of Memex as an "enlarged intimate supplement to memory" (1945), Weiser's vision of ubiquitous computing where computers "vanish into the background" (1991) and Ishii and colleagues' vision of three-dimensional interfaces that are "as reconfigurable as pixels on a screen" (2012). At a given time, HCI is driven by a multitude of research visions and it is possible to examine the same interface or technology through the lens of different visions. Two research visions, however, can also lead to conflicting design outcomes; for example, direct manipulation interfaces and intelligent agents presented competing visions for what makes an ideal user interface, which led to debates about their respective merits (Maes et al., 1997). The current interest in BCIs similarly seems to conflict with some other HCI visions such as the vision of tangible computing that emphasizes a rich repertoire of physical gestures. And as with other technologies, there is more than a single way to frame the future utility of BCIs. Here, we find one potential framing particularly relevant for embodiment. We identify this as the vision of providing more direct access to users' mental states than allowed by existing input devices that require users to perform body movements.

We observe that while many BCI researchers (e.g., Farwell & Donchin, 1988; Spapé et al., 2015; Tan & Nijholt, 2010; Wolpaw et al., 2000) caution against the hype of "wire-tapping" or "mind reading" interfaces, the claims of directness still feature strongly in some future visions of BCIs. Schalk identifies the conventional interface between humans and computers as an impediment to future human-machine symbiosis and argues for the "theoretical and practical possibility that direct communication between the brain and the computer can be used to overcome this impediment by improving or augmenting conventional forms of human communication." (2008). Along the same lines, Ebrahimi and colleagues identify conventional interfaces as the "weak link in communication" and ask "Are there ways to altogether bypass the natural interfaces of a human user such as his muscles and yet establish a meaningful communication between man and machine?" (2003). This is echoed by Krepki and colleagues: "Since all these information streams pass its own interface (hand/ skin, eye, ear, nose, muscles) yet indirectly converge or emerge in the brain, the investigation of a direct communication channel between the application and the human brain should be of high interest to multimedia researchers" (2007). A related promise is direct communication between multiple brains through EEG and TMS (transcranial magnetic stimulation): "Until recently, the exchange of communication between minds or brains of different individuals has been supported and constrained by the sensorial and motor arsenals of our body. However, there is now the possibility of a new era in which brains will dialogue in a more direct way" (Grau et al., 2014).

Our aim in presenting these statements is not to generalize over BCI research or to suggest that bypassing the body exclusively accounts for the potential benefits of BCIs. We rather find the statements positively thought-provoking as they point to an actual challenge faced by HCI, namely determining the relevance of the wider body for human–computer interaction in a potential era of viable BCIs. We also note that statements such as those cited above reflect an understanding of the human body that has been influential in HCI long before the advent of BCIs, namely as a limited capacity motor channel between the mind and the computer (Card et al., 1983). Once the body beyond the brain is conceived as a communicational bottleneck, the promise of bypassing it has immediate appeal (Figure 1). For us, it is this intellectual parallel that makes the research vision worthy of further examination.

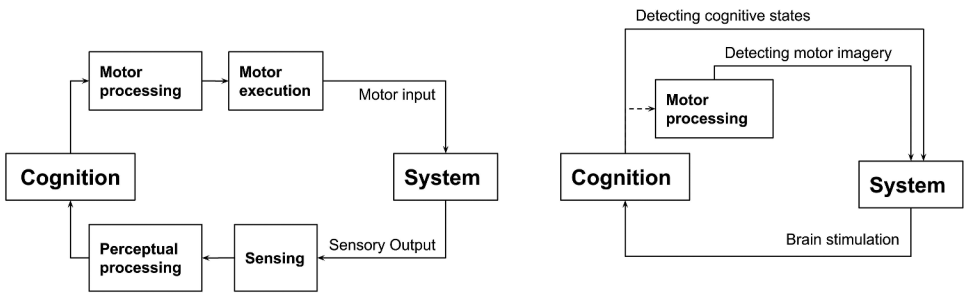

**Figure 1.** A vastly simplified model of interaction based on an understanding of cognition and motor execution as a staged process (after Card et al., 1983). In comparison to motor input (left), BCIs (right) allow bypassing body movements. Note that directness can come in different degrees (e.g., directly accessing thoughts or sensing motor imagery).

Many shortcomings of this conception of the body have since been highlighted. Before moving into criticism, however, it is useful to formulate a strong version of the research vision by laying down a more qualified understanding of directness. Relevant concepts can be found in HCI work that treats human–computer interaction in terms of information transmission between the mind and an interactive system. We particularly turn to "expressiveness" and "effectiveness" criteria (Card et al., 1991) that provide a comprehensive list of considerations for input device evaluation.

## 2.2. *Directness as information transmission criteria*

Directness can first be understood in terms of more "expressive" (Card et al., 1991) interfaces whose semantic space better matches the prior mental content. Interface evaluations often make the idealized assumption that the semantics of an application program adequately represent the mental content to be communicated. Yet one can also make the case that existing input devices provide a restricted vocabulary for expression. One motivation for BCI research has been the perception that brain signals provide a more accurate representation of mental content than allowed by other input methods (Spapé et al., 2015). As others noted (Fairclough & Gilleade, 2014), however, the current state-of-the-art falls short of this ideal; except for a limited number of generative interfaces (e.g., Kangassalo et al., 2020), BCIs often reduce rich brain signal data into a limited set of input actions. An interface should additionally be able to communicate only the intended meaning and nothing else (Card et al., 1991). Challenges might emerge for future BCIs when brain activity – so far reserved for thinking – is used for interaction. In eye tracking research a similar challenge has long been known as the "Midas Touch" (Jacob, 1990) problem; a user can gaze at a certain location on the interface to perceive information, but the use of eye movements as an input can cause unintended commands. There seems to be considerable overlap between the neural activation patterns of imagination, observation, and execution of motor actions (Decety, 1996; Grèzes & Decety, 2001). While differentiating motor execution from its imagination can still be possible, for able-bodied users this might raise the challenge of collecting brain signals for interaction without having to perform the actual motor execution.

Assuming that a message can be expressed through the interface, there remains the challenge of doing this *effectively*, which can be understood in terms of task completion time, ergonomy or any other criteria (Card et al., 1991). Completion time, a metric that input devices are often evaluated against, is a combination of multiple factors such as latency and bandwidth. In HCI, the time needed for acquiring an input device, such as by reaching a mouse, approximates to latency. When the body is conceptualized as a communication channel, the scope of latency extends to the lag *within the body*. Among other factors, this is determined by the conduction velocity of the peripheral nerves, which can be calculated by observing the delay between cortical stimulation in the brain and the electrical activity of muscle tissue (Rothwell et al., 1991; Stetson et al., 1992). One motivation for using BCIs has thus been the elimination of this inherent lag in peripheral motor control. For example, a BCI can detect when a car driver is about to reach for the brake pedal in an emergency, and brake the car before actuating the foot pedal (Haufe et al., 2011).

Another effectiveness metric is the bandwidth of a channel. Yet modeling body parts in terms of bandwidth has been non-trivial. Unlike the communication bandwidth of machines that can be specified by design, abstraction of the human body in terms of bandwidth relies on the throughput that is observed in natural or experimental settings. There exist several metrics for measuring BCI throughput. Information transmission rate (Wolpaw et al., 2000) combines classification accuracy of brain signals with timing information, while metrics such as efficiency (Bianchi et al., 2007) and utility (Dal Seno et al., 2010) also take error recovery into account. Task-specific metrics such as text entry speed enables direct comparison between BCIs and manual input devices. Text entry speeds of everyday typists on a physical keyboard can range between 34–79 words per minute (wpm) (Feit et al., 2016). Intracortical BCIs inserted near the brain's motor cortex can reach close to 8 wpm using a single pointer for character selection (Pandarinath et al., 2017). Recently, linguistic content was communicated over electrocorticogram (ECoG) signals that were recorded while subjects were speaking during an experiment (Makin et al., 2020). This resulted in high throughput but came with the major pitfall of requiring the subjects to read aloud the sentences. The post hoc analysis accordingly showed that the brain regions that contributed most to the prediction were those associated with speech production. Such intracortical BCIs are invasive and thus reserved for disabled users. For noninvasive BCIs, signal quality and typing performance have so far been lower (Townsend et al., 2010). Many BCIs accordingly compensate for the limited throughput by using time-multiplexing, for instance by using event-related signals to select letters from a matrix while rows and columns are sequentially highlighted (Farwell & Donchin, 1988).

We so far treated the selection of input techniques as an exclusive problem. However, the parallel use of multiple sensing devices can increase the overall throughput provided that the data streams provide independent information. Information that is gathered from different modalities is redundant when BCI signals simply depend on muscle activity; decoding text from ECoG while speaking (Makin et al., 2020) might not provide additional information in comparison to what can be sensed using a microphone. The long-term goal of using BCIs for healthy individuals is to sense what is not easily observable through the wider body (Allison et al., 2007; Ebrahimi et al., 2003; Fairclough, 2011) and filter out artifacts that result from synchronous muscle activity, either by restricting body movements or processing data after the experiment (Fatourechi et al., 2007; Wolpaw et al., 2000).

### 2.3. *Summary*

The treatment of interaction as information transmission between the mind and the system and the related conception of the body as a communication channel leads to questions such as:

- How expressively does an interface communicate prior mental content?
- How to avoid unintended interface commands?
- What is the latency and throughput of different input methods?
- How much effort/training do different input methods require?
- What are the interferences and redundancies between different input methods?

This formulation of interaction is often a necessary simplification to deal with the complexity of human behavior, and information transmission criteria indeed provide a list of considerations that is useful to understand potential benefits and practical challenges ahead. Current technological limitations also point to alternative use cases in which BCIs do not completely replace body movements but are used alongside them, for instance, to infer mental load during manual interaction (Afergan et al., 2014; Yuksel et al., 2016) or detect user errors (Vi et al., 2014). Some argued that such "passive" BCIs, which use brain activity that occurs as part of another activity, could more readily benefit able-bodied users when compared to real-time control interfaces (Erp van et al., 2012).

A more fundamental problem is the conception of the human motor system as an information channel, which has been beset by difficulties since the early applications of information theory to the human body.[1] Consider Fitts' Law which conceptualizes pointing performance in terms of the information capacity of the human motor system. While the law serves well as a statistical tool for predicting pointing performance, defining it in information-theoretic terms of source, encoder and channel has been problematic (Gori et al., 2018; Seow, 2005). Some (e.g., Gori et al., 2018; Zhai et al., 2012) specify the user intention as the source and motor mapping as the encoder but avoid grounding these constructs to actual physical entities in the human body. A difficulty facing this task is defining an appropriate boundary between a putative source and an encoder, which would require researcher-defined functional constructs to have direct corresponding entities within the body. The evidence from embodied cognition showed some of the problems with this endeavor. The next section summarizes the main discussion points.

## 3. Cognition and the body

This section presents the main claims of embodiment concerning cognition and the body by contrasting them to what has been later referred to as cognitivism.[2] While a thorough review is beyond our scope, a quick overview is necessary to lay the groundwork for the present discussion. In what follows, we summarize the embodied claims regarding 1) the status of mental representations, 2) the relationship between meaning and interaction, 3) the neural basis of cognition, and 4) the extent cognition is integrated with the body and the wider environment. We discuss the relevance of these claims for the present discussion and identify a modular understanding of cognition, related to the fourth discussion point, as the commitment that is most relevant for the vision of providing more direct access to mental states.

### 3.1. *The status of mental representations*

For cognitivism, cognition was primarily the transformation of mental representations (Fodor, 1975). The precise meaning of the concept "representation" is debated (for a discussion see Haselager et al., 2003), but in most cognitive science the term has come to mean distinct inner states that stand for external or imaginary things such as physical entities, plans, memories or counterfactuals. Mental representations allow thinking in an off-line fashion, namely in the absence of activity-relevant interaction with the environment. When deployed to explain human problem-solving, the human ability for such off-line cognition led some to conceptualize mental representations as analogous to computer programs in that they adequately specify the sequence of operations needed for task completion (Miller et al., 1960) or as adequate descriptions of a desired end state, which consequently motivated the long line of HCI research on plan and intent recognition (e.g., Allen, 1999; Horvitz et al., 1998).

Embodied approaches scrutinize the assumption of well-formulated mental representations that underlie human behavior. Human behavior can instead be explained in terms of habitual responses to environmental stimuli that bring an agent closer to a situation (Dreyfus, 2002) and relies on the presence of many external conditions that have no mental counterpart at all (Suchman, 1987)—even though it is always possible to attribute intentions or plans to an agent in retrospect. The criticism against the need for explicit plans or goals is not an all-around rejection of mental representations, but it significantly limits the extent they account for human competence. From an embodied viewpoint, mental representations can be more generally characterized as being in coordination with the environment as opposed to representing its current or preferred future state.

---

[1]In fact, HCI lacks a uniform mapping of information-theoretic constructs. BCI research often treats the interface as the information channel and the background brain activity as the noise (Kronegg et al., 2005). Research on muscle-operated input devices, in contrast, generally treat the motor system as the channel.

[2]Both embodiment and cognitivism are better described as umbrella terms. Cognitivism generally modeled human cognition in terms of mental information processing, but what constitutes its core has been the subject of debate, with researchers alternatively emphasizing the assumption of abstract symbolic processing (Clark & Chalmers, 1998; Vera & Simon, 1993) or a strict internal–external distinction (Agre, 1993). Embodied perspectives commonly object to the commitments of cognitivism, but include a variety of different approaches that are grounded in phenomenology and revised views in cognitive science (Marshall & Hornecker, 2013). Work outside HCI cataloged different meanings the term embodiment can stand for (Anderson, 2003; Rohrer, 2007) and there have been calls to treat embodied cognition as individual claims (M. Wilson, 2002).

The vagueness of mental representations accepted, the promise of interfaces that allow "mind-reading" or "wire-tapping" to intentions may lose some of its appeal as the main challenge in interaction does not always lie in the execution of a well-formulated goal. Does this insight completely diminish the value of more directly accessing mental states? Two reasons make us argue against this. First, the motivation for accessing mental states does not necessarily imply the instant execution of an end goal; it can instead be formulated in terms of facilitating step-by-step interaction in a better way. For instance, a BCI does not need to "mind-read" and transcribe a text phrase at once; it can allow character-by-character entry to facilitate word suggestion, but in a way is ideally more efficient than manual input. Second, mental states do not necessarily correspond to intentions but can rather stand for any data that can be used by a system to generate an appropriate response. A case in point is interfaces that target the implicit use of brain signals. A search interface, for instance, can use brain signals to infer user interests (Eugster et al., 2016). In doing so, it can aim to decrease the mental effort of query formulation rather than the physical effort of text entry. We thus find this commitment only partially relevant to the research vision of directly accessing mental states.

### 3.2. *The relationship between meaning and interaction*

The emphasis on mental representations also supported the methodological choice of conceptualizing the value of an interactive process, that is its *meaning*, in terms of users' prior intentions. A practical challenge for this choice is that users' mental states are not accessible to researchers in the way computer states are. In both cognitive science and HCI, researchers addressed the challenge by formulating activity in terms of tasks that stand as a proxy for user intentions and assuming rational behavior on users' side (Card et al., 1983; Simon & Newell, 1971)—in a way defining the meaning of user actions prior to and independent from interaction. The researcher's privileged position in framing the activity in terms of tasks and setting the metrics for evaluation is foremost a methodological preference, but also one that is reinforced by the decision to treat human-computer interaction as a "microworld" (Newell & Card, 1985) that is studied in isolation.

For agents situated in daily contexts, however, the significance of interaction often depends on self-defined criteria that are intelligible to the agents themselves. Thus, some in HCI expanded the scope of embodiment to cover a methodological commitment toward identifying how meaning is created by users as part of their ongoing interaction (e.g., Dourish, 2004; Robertson, 2002). A methodological implication of this commitment has been to acquire thick descriptions of a situation instead of abstracting activities a priori into tasks and metrics (Dourish, 2004; Harrison et al., 2007; Klemmer et al., 2006). In parallel, the scope of research expanded beyond the micro-world of human–computer interaction to include externalities such as organizational behavior and social cognition. Aspects of public availability and interpretation came to the fore, one typical example being collocated interactions where users' arm and head movements become visible to each other (Robertson, 2002).

Future BCIs can reduce the information available to other agents if they eliminate the need for body movements.[3] Yet they can also create new opportunities for social interaction if brain signals themselves are made available to others (e.g., Hassib et al., 2017). It should, however, be kept in mind that different approaches to meaning result in diverging implications for research. A thorough comparison has been made by Boehner and colleagues in the context of physiological computing and emotions (2007); whether meaning is treated as prior to or emergent during interaction can lead to different design decisions concerning the extent physiological data is abstracted into predefined categories by the system or left to users' own interpretation. In a task-based approach, a message composed by a BCI or a manual typing interface can be evaluated solely on external measures of performance. In the expanded embodied view, however, the evaluation between different interfaces is not independent of how their qualities are interpreted by other users. A BCI-composed message can alternatively be interpreted as more authentic, lazy, or ambiguous depending on the receiver.

---

[3]Note that, this is not the case when the technical shortcomings of BCIs indirectly facilitate similar visibility by requiring users to perform body movements (O'Hara et al., 2011).

### 3.3. *The neural basis of cognition*

Early models in cognitive science also conceived cognition as a generalized and amodal processor, whose functioning rules are independent of the physiology of the body (Fodor, 1983; Vera & Simon, 1993). Cognition has instead been modeled after the computer architectures of the time, based on symbol-manipulation (Fodor, 1975; Simon & Newell, 1971), sequential processing, and the separation between processing and memory functions (Atkinson & Shiffrin, 1968; Sternberg, 1969)—conventions that are followed by the Model Human Processor in HCI (Card et al., 1983). The early expectation was that neurophysiology would eventually catch up with the abstract, symbolic models (Simon & Newell, 1971, p. 158).

Needless to say, the progress has also been in the other direction; researchers developed biologically-informed models that do not separate memory and processing (e.g., connectionist models (Rumelhart & McClelland, 1986)) to explain lower-level human abilities such as visual recognition. The disciplinary tensions can be observed within the "imagery debates" concerning whether mental phenomena should be modeled as "propositional" (i.e., amodal and symbolic) or "depictive" (i.e., modal and preserving the spatial properties of the sensorimotor objects through the topographic organization of the brain) (Kosslyn, 1994). The latter view ultimately gained prevalence through multiple lines of evidence (M. Wilson, 2002).

What is the relevance of the neural basis of cognition for the present discussion? It can be argued that when the brain is interfaced, its physiology and the localization of neural activity come to the fore in a way they do not with the abstracted models of cognition. In this sense, the motivation for more direct access to mental states does not explicitly assume an abstract model of cognition. At the same time, while not assumed, whether cognition is treated as abstract or embodied has implications for what we understand as mental states. For instance, the word "decoding" is often used in the context of BCIs (e.g., Rao et al., 2014; Schalk, 2008), but its operationalization depends on one's research commitment. The concept of decoding involves representing an already symbolic structure in the mind for abstract models of cognition that assume the independence of mental states from the physical structures of the brain. In embodied accounts, however, mental states refer to individual and modality-specific neural activities whose decoding into information depends on the particular method employed.

### 3.4. *The extent cognition is integrated*

Closely related to the neural basis of cognition is the extent cognition is integrated with sensorimotor processes and the environment. What exact properties define integration has been the subject of analysis within embodied cognition literature (Clark & Chalmers, 1998; Heersmink, 2015; Kirsh, 2019) and some relevant properties are "coupling", "persistence" and "individualization". Relatively loose coupling (i.e., consisting of unidirectional and low-throughput interactions) and limited individualization between cognition and sensorimotor processing characterize cognitivist models; rich sensorimotor data is abstracted into amodal and symbolic information before cognitive processing and vice versa (Fodor, 1983). This leads to a more or less modular system in which cognition is sharply separated from sensorimotor processing. The sharp boundaries between cognition and sensorimotor processing, by extension, exclude interaction with the environment from the scope of cognition, leading to a strong internal–external distinction.

Modality-preserving models of neural representation in embodied views, by definition, assume individualized and highly coupled relationships (i.e., consisting of bidirectional, high-throughput interactions) between different physical structures that come into play during cognition. Yet there have been multiple claims regarding the scope of integration.

The first, internalist, claim is the integration between sensorimotor processes and thinking *inside the brain*. This partly refers to the real-time effect of sensorimotor variables (e.g., moving one's hand toward or away from the body) on seemingly abstract activities such as language cognition, memory retrieval or problem-solving (Barsalou et al., 2003; Dijkstra et al., 2007; Glenberg & Kaschak, 2002; N. L. Wilson & Gibbs, 2007; Thomas & Lleras, 2009). Furthermore, body and tool use seems to have causal-historical roles in individualizing brain structures as seen in the metaphorical mappings between abstract concepts and physical experiences in language (Lakoff & Johnson, 2008). In HCI, some work directly builds on this line of research and utilizes concepts such as image schemas (Hurtienne & Israel, 2007) and embodied metaphors (Antle et al., 2009) to design interface controls in a way that preserves their control variables' mappings to the body. The representation of the body shape and posture in the brain, the "body schema" (Berlucchi & Agliotti, 1997; Maravita & Iriki, 2004), is another example of individualization as is the phenomenon of "tool extension", namely the changes to the body schema through the use of tools (Berti & Frassinetti, 2000; Iriki et al., 1996).

The second, externalist, claim expands cognition's unit of analysis to encompass what happens *outside the brain*. Already in the 1970s, the ecological approach argued that the world itself provides an abundance of information for action, namely affordances, which obviates the need for extensive mental representations (Gibson, 1977). Extended views of cognition echo this view and call for recognizing the external world as part of a larger cognitive whole (Clark & Chalmers, 1998; Clark, 2007). To the HCI community, this position is best known through the research program of "distributed cognition" (Hollan et al., 2000). Distributed cognition attributes some of the cognitive science constructs such as information, memory, and representation—that are traditionally reserved for mental phenomena—to the environment, in a way blurring the previously sharp boundaries between internal processing and external interactions (Hutchins, 1995). Thinking, eye movements, and other motor actions instead become interchangeable methods for information processing. People can eliminate mental rotation or translation activities by manipulating the objects in the environment (Kirsh & Maglio, 1994), ease perception by hiding or cueing certain features, or use hands as placeholders to decrease mental memory load (Kirsh, 1995). As such, there is less emphasis on the communicative bottleneck between the mind and the environment, and an increased sensibility toward how cognitive processes are offloaded onto the environment. This consequently leads to a less persistent model of cognition that is partly composed of temporary interactions with external structures.

Whether the scope of cognition should be delimited to the mind, to the body, or even to the body-plus-environment is a matter of definition and the epistemic trade-offs that come with different units of analysis have been discussed elsewhere (M. Wilson, 2002). The degree of modularity or integration between different entities, however, is an empirical question with practical consequences. Here, we argue that the strict separation that modularity entails is central to the motivation of bypassing body movements; when central cognition is conceived as prior to and independent from body movements, the motor system can ideally be bypassed for direct access. We thus view the extent cognition is integrated as the central concern for research.

### 3.5. *Summary*

We outlined the different commitments of cognitivist and embodied approaches and assessed which commitment provides a more explicit conceptual basis to the research vision of providing more direct access to mental states. The *extent mental representations are adequate descriptions* for action is only partially relevant to the present discussion as the benefits of BCIs do not depend on prior intentions providing the satisfaction conditions of interaction. The conception of *cognition as abstract information processing* is also not explicitly assumed by the research vision, but what is understood as mental states (i.e., the extent they are modality- and physiology-specific) has implications for what it means to decode them. The *relationship between meaning and interaction* is less specific to the present discussion but has general methodological relevance. A *modular view of cognition* is the commitment that we argued to be the most relevant for the motivation of bypassing the body and providing more direct access to mental states.

## 4. Implications of an integrated view

The utility of an integrated conception ultimately relies on its ability to guide research in the form of generating research questions, explanations, predictions, and interface solutions that are not easily conceivable from a modular understanding of cognition. In this section, we show how embodiment can lead to future research directions by referring to previous studies, which we selected to showcase the integration between sensorimotor and other cognitive processes in a variety of domains. We then discuss the more general implications for methodology.

### 4.1. *Functional relationship between body representations, motor imagery, and BCI performance*

It has been shown that motor imagery in the brain can be accompanied by electromyographic (EMG) signals on the arm (Guillot et al., 2007) and face (e.g., during internal verbalization of speech; see Kapur et al., 2018; Nalborczyk et al., 2020), one possible explanation being the incomplete inhibition of peripheral neural activity during motor imagery (Guillot et al., 2007). These peripheral correlates to motor imagery should be distinguished from the more radical embodiment claim of tight interconnections between sensorimotor processes and conceptual knowledge, but they support the claim of shared neural substrates between motor imagery, action observation and execution (Grèzes & Decety, 2001; Jeannerod, 1994; Rizzolatti et al., 1996). Other observations point to situations in which the actual execution of body movements both reflects and facilitates internal cognitive processes. For example, studies documented spatial overlaps between gaze patterns and the layout of an imagined visual scene (Brandt & Stark, 1997; X. Wang et al., 2019), a finding that has been explained through eye movements' role in retrieving and processing visual information in the absence of sensory stimuli (Ferreira et al., 2008). There is also strong evidence for the memory-facilitating role of hand gesturing during speech, in addition to its communicative function (for a review see Pouw et al., 2014). Interestingly, such gesturing rarely occurs during thinking, hinting at gestures' role in mitigating the effect of increased cognitive effort expended during speech production (Hostetter & Alibali, 2008).

The overlaps between motor planning and execution in the brain would also predict BCI performance to be partly dependent on motor skills. In line with the expectation, users' fine motor manipulation skills (Hammer et al., 2012), handedness (Zapała et al., 2020) and daily amount of hand and arm movements (Randolph et al., 2010) have been shown to account for some of the individual variance in motor imagery-based BCI performance. Along the same lines, BCI performance should be lower in the absence of motor ability. Patients in complete locked-in-state indeed show degraded BCI performance, but this finding does not hold for partially paralyzed (Kübler & Birbaumer, 2008) highlighting the need for more research to uncover the mechanisms that underlie both BCI and motor performance. Besides motor imagery, motor skills seem to affect brain activation during language comprehension. Hockey players have been reported to show greater activation in the left premotor cortex compared to non-hockey players when presented with sentences that describe hockey actions (Beilock et al., 2008).

### 4.1.1. *Research direction: Measuring the effects of motor variables on BCI performance*

Numerous studies have recorded brain signal data while experimental subjects perform motor activities such as lifting/grasping (Agashe et al., 2015;Luciw et al., 2014) or walking (Gwin et al., 2010; Kline et al., 2015). Part of the motivation for this line of work is improving the usability of brain recordings in naturalistic settings (e.g., BCI use when walking) by identifying and removing signal noise related to body movements (Gwin et al., 2010; Kline et al., 2015). Yet little work has measured the effect of task-relevant motor variables, a research gap noted within the brain imaging community (Gramann et al., 2014). Thus, a potential research direction is evaluating the effects of task-relevant motor variables on task and BCI control performances. Researchers can, for instance, observe whether gesturing can improve BCI performance through its effect on mental imagery, in the same way it aids speech production (Pouw et al., 2014).

### 4.1.2. *Research direction: Testing the relationship between motor skill and BCI performance*

HCI research can also extend the previous work (e.g., Randolph et al., 2010) that aims to find correlates between motor skill in using input devices and BCI performance. While previous work has investigated the role of naturalistic motor skills, HCI researchers can study the effect of skills related to input device use, for instance, by testing if expert gamers have a performance advantage for certain types of BCIs. For motor imagery-based interfaces, researchers can measure the motor skills of experimental participants when using different input devices (e.g., using a stylus for drawing shapes or moving objects on a touchscreen) and test if these correlate to performance when operating BCIs that use similar motor imagery.

Other experimental designs can test whether motor skill advantage extends into reactive BCIs that use event-related potentials (ERPs). ERP-based BCIs often employ symbols (such as letters) for eliciting brain signal responses. Employing more complex visual stimuli that refer to motor capabilities can enable investigating correlations between motor skill and ERP-based BCI performance. For example, we referred to hockey players' increased brain activation when shown language stimuli (Beilock et al., 2008). This observation would predict better performance for hockey players when shown visual stimuli such as a hockey stick.

Finally, researchers can test whether BCI performance can be improved by training users with non-BCI interfaces that rely on similar mental imagery, by observing performance differences between the trained and untrained groups.

### 4.2. *Control mapping and compatibility in perception and action*

The integration within the cognitive system would also predict BCI performance to positively correlate with motor variables that are congruent with the control task. Performance studies have long observed that control – display similarity, in terms of spatial and nonspatial factors, results in higher performance for visuoperceptual motor tasks (Fitts & Seeger, 1953), a finding that also holds for BCIs (Thurlings et al., 2012). The phenomenon has previously been explained through information processing models and the formal analysis of input devices (Card et al., 1991; Fitts & Seeger, 1953). Recent studies aim to find neural correlates to performance differences and demonstrate the effect of more complex sensorimotor variables. Congruent tactile stimulation on subjects' hands during motor imagery tasks seems to improve BCI accuracy (Shu et al., 2017). Studies on tennis (Bisio et al., 2014;Mizuguchi et al., 2015) and badminton (Z. Wang et al., 2014) players show that holding a racket improves motor imagery, with effects being dependent on player experience and hand posture (Mizuguchi et al., 2015).

### 4.2.1. *Research direction: Tangible implements to support motor imagery*

The observations related to the role of manual artifacts in supporting motor imagery also point to potential benefits that can result from incorporating tangibles into BCI use. Extending previous work in motor imagery research (Bisio et al., 2014; Mizuguchi et al., 2015; Z. Wang et al., 2014), the field of HCI can draw on its design and fabrication capabilities to create physical artifacts that support BCI performance. For motor imagery-based BCIs, these artifacts can aid users in assuming the posture related to an imagined movement and compare the performance to a control condition in which the tangible artifact is not present. An example experimental setup can instruct participants to imagine hand grasping movements in the presence or absence of a tangible that facilitates grasping posture, and compare the BCI performance of the two conditions.

### 4.2.2. *Research direction: Embodied constraints for the brain*

Another research direction that builds on control mapping congruence is the application of "embodied constraints" to BCIs. In HCI, embodied constraints refer to task-specific constraints that are encoded directly into the environment as a way to decrease the cognitive burden on users (Dijk, 2009; Hornecker & Buur, 2006). Consider the example of a tangible urban-planning interface (Underkoffler & Ishii, 1999), the impossibility of interpenetrating the tangible tokens that represent buildings is a desired constraint as it mirrors the impossibility of the actual task. One advantage of having such task-specific information inside the environment, as opposed to the brain, is the convenient configurability of physical structures. Unlike mental representations that form through practice, physical structures can be easily replaced depending on the task at hand. At the same time, recent work shows how such embodied constraints can move closer to the body with the help of electromuscle stimulation, for example, by constraining users' approach to tangible objects before any physical contact (Lopes et al., 2015).

BCI equivalent of embodied constraints would similarly impose task-specific constraints within the brain based on the task at hand. While cortical interventions might not be technically possible and safe in the near future, the interdependence between sensorimotor processes and brain activity points to alternative methods for achieving similar effects. For example, we noted the effect of sensorimotor variables such as hand posture (Mizuguchi et al., 2015) or tactile stimulation (Shu et al., 2017) on motor imagery performance. Consider how these observations can be synthesized with the design insight of embodied constraints: Based on the requirements of a task, an interface can reinforce or suppress certain mental imagery by changing visual, tactile, or somatosensory stimuli (e.g., by modifying users' hand or body posture using an exoskeleton). To illustrate, a system can give tactile feedback on the right-hand side of the body when navigating left is not possible in a virtual environment (to match the actual spatial constraint of the environment). No such interactive system exists to our knowledge, but one can recognize how it represents an alternative to the ideal of direct mental access; with embodied constraints, the design focus shifts from the accurate sensing of prior mental content to reinforcing task-related neural activity through stimuli.

### 4.2.3. *Research direction: Body-based metaphorical mappings*

Finally, the metaphorical mappings between the body and abstract concepts would predict the interface controls that are congruent with these mappings to outperform design alternatives that are incongruent. For example, phrases "feeling up" or "feeling down" describe sentiments as well as physical directions (Barsalou et al., 2003; Lakoff & Johnson, 2008) and we would expect these sentiments to be congruent respectively with imagining upward or downward movements. Future studies can test this prediction by observing whether the design of BCI actions, such as leaving an emotional reaction on social media, would benefit from similar mappings; an experimental setup can compare the performance and satisfaction of congruent (positive emotion mapped to upwards movements and vice versa) and incongruent motor imagery for leaving emotional reactions.

### 4.3. *Modality-dependence of language*

Language production is a common daily task and various modalities such as speaking or writing often lead to different language expressions. The difference can be partly explained through the distinct communicative strategies used in oral or written contexts but also through the particular sensory and motor demands of each modality; writing and speaking utilize different muscle sets and sensory feedback, which leads to different cognitive constraints on performance as well as different brain activity patterns (Brownsett & Wise, 2009). Yet the extent these sensory and motor processes are integrated with semantic and lexical processes (i.e., processes related to meaning and word selection) has been the subject of debate (for reviews, see Chatterjee, 2010; Meteyard et al., 2012). Some models of language production (e.g.,Dell, 1990; Roelofs, 1992) conceptualize these processes as modality-neutral while others (e.g., Caramazza & Miozzo, 1997) propose modality-specific models. The extent of integration, and thus embodiment, is relevant as modality-specific conceptions of language lead to different evaluation goals than studies that assume the independence of the semantic and lexical stages from motor articulation.

#### 4.3.1. *Research direction: Observing qualitative differences between BCI-based language production*

HCI studies generally assume a modality-neutral model as they operationalize text entry in terms of externalizing prior thought and compare them based on their effectiveness; subjects are tasked to transcribe a memorized fixed phrase with a given method as quickly and as error-free as possible. If different methods for language production are modality-specific, however, evaluation should observe distinct linguistic outcomes that result from different interfaces. Consider different types of language production that can be facilitated through BCIs:

- A BCI that reproduces writing experience by relying on the motor imagery of handwriting movements, as in (Willett et al., 2021),
- A BCI that reproduces speaking experience by relying on the motor imagery of speech, as in (Makin et al., 2020).

An integrated model of language production would regard the two BCIs as facilitators of distinct lexical selection processes in the brain instead of different outlets for the same lexical content as the earlier semantic and lexical stages are not assumed to be independent of motor articulation. We would then expect their evaluation to observe not only the text entry speed but also the qualitative differences in lexical selection and language production. While current BCI prototypes reproduce existing methods of language production (e.g., speech, handwriting), we might witness the emergence of novel BCI-specific forms of text entry that bypass lexical selection altogether, for instance, by generating text from user reactions to verbal or visual stimuli. The variance in language output becomes even more relevant for such interfaces and the ability to provide appropriate text recommendations becomes a more applicable evaluation criterion.

### 4.4. *Affect and emotion in the body and the brain*

Emotion research has received much attention in BCI and physiological computing (Beale & Peter, 2008; Fairclough, 2009). Going beyond mere detection of emotions, affective computing as an area has addressed how interacting with computers can influence emotional states and how emotional expression can be synthesized in virtual agents (Tao & Tan, 2005; Torres et al., 2020). Despite such successes, researchers have not reached a consensus on a unified theory of how the behavioral correlates and physiological expressions defining emotions are related to psychological feelings and subjective experience, or affect. Historically, embodied theories of emotion tend toward the so-called James-Lange theory of emotion (Prinz, 2008), which hypothesizes that affect is caused by physiological states and behavioral changes as opposed to cognitivist theories of emotion giving cognition and affect a primary role in causing behavior (Oatley & Johnson Laird, 2014).

Thus, more modern advocates of the theory suggest that perceiving and thinking about emotion involve "perceptual, somatovisceral, and motoric reexperiencing (collectively referred to as 'embodiment') of the relevant emotion in one's self" (Niedenthal, 2007). Despite the lack of unified theory, however, recent reviews demonstrate that emotion detection is most effective when using the widest variety of physiological and behavioral signals available, for example from video, audio, other content, central or peripheral physiology, gaze, and kinesthetic expression (Ahmed et al., 2020). Thus, computer techniques and applications in this area are useful as practical demonstrations of how information about emotion can be extracted from brain signals, other physiology or embodied behavior, and may thereby provide data toward theory unification.

### 4.4.1. *Research direction: Further develop multimodal approaches to emotion combining embodied aspects with brain signals*

BCI research has long concentrated on brain signals, for obvious reasons, but embodied frameworks suggest a stronger role of peripheral and behavioral signals in determining affective states. Thus, multimodal signals, such as electrocardiography (Ferdinando et al., 2016), facial electromyography (Fridlund & Cacioppo, 1986), electrodermal activity (Feng et al., 2018), and other more peripheral psychophysiological signals may provide complementary information to BCIs, particularly with regards to detecting affective states. Furthermore, in parallel to what is considered for other BCI tasks, the research can include more naturalistic settings where body movements are not constrained and can be effectively detected given that they might play a role in perceiving or thinking of emotions. Given advances in the availability of remote sensing, future research must establish how effective a multimodal, embodied approach relying on rich data combining situational, behavioral, physiological, and neural data can be, not least to mitigate near-future ethical risks (Steinert & Friedrich, 2020).

### 4.4.2. *Research direction: Investigate communicating affect and emotion using BCI*

Emotions play an important role in interpersonal communication and will become more important to HCI in general, for example with regard to mediated and human–AI interaction. Thus, BCIs may provide emotionally augmented communication, providing complementary information to current interfaces. First, passive BCIs detecting emotions can provide an affective subtext to inform interactants of the intended meaning behind, or affective consequence of communicative acts (Harjunen et al., 2018; Kuber & Wright, 2013; Spapé et al., 2019). Both active and passive BCIs may furthermore enable emotions by themselves to become a language of interaction. For example, by using neurofeedback disposition and arousal toward avatars can be learned to be controlled (Cavazza et al., 2014). Similarly, implicit detection of negative affect may be used to adapt a user interface's complexity and improve engagement (Aricò et al., 2017; Lawrence et al., 1995).

### 4.5. *Summary and methodological implications*

We described various ways embodiment and an integrated understanding of cognition can guide research. These include providing body-grounded explanations for BCI performance, designing BCIs in a way that is compatible with existing body mappings, and proposing evaluation considerations that are neglected in modular views of cognition. Some research directions build on evidence from embodied cognition while others extend embodied interaction design heuristics into the BCI domain. Below, we describe the more general methodological implications of the above-described research directions.

### 4.5.1. *A clearer rationale for BCI applications*

The many ways in which cognition in the brain is dependent on and mirrored across the body should motivate researchers to clearly state their rationale for BCI applications targeted toward able-bodied users. In particular, we propose an analytic distinction between using BCIs as 1) a source of neurological data and 2) an enabler of interaction without motor execution.

Firstly, observations related to offloading cognitive activity onto the wider body (as in gesturing during speech) and improved motor imagery in the presence of compatible body postures (e.g., Bisio et al., 2014; Mizuguchi et al., 2015) suggest that BCI usage for the general population will coexist with and benefit from concurrent motor execution. As such, researchers should be ready to demonstrate the rationale of BCIs *in the presence of motor execution*, for instance, by observing whether brain signals provide useful information in addition to what can be sensed from body movements.

Conversely, eliminating motor execution should not automatically justify BCI usage as brain activity can be mirrored across the body even in the absence of body movements (Guillot et al., 2012) and can closely reflect the environmental stimuli. As BCI research partly moves from lab-conducted studies and disabled users into rich everyday settings and the general population, useful information related to neurological states can partly be inferred through sensing the environment, eye movements, or peripheral nerve activity that occurs without motor execution (e.g., using electromyography). It is then necessary to establish the rationale of using BCIs *even when motor execution is purposefully eliminated*. For example, researchers can record brain signals, peripheral nerve activity and environmental stimuli in synchrony, and compare their success in predicting what users consider appropriate system behavior.

### 4.5.2. *Caution when restricting body movements*

The analytical distinction made above also calls for exercising caution when restricting body movements in order to reduce "data artifacts" (i.e., undesirable signals that interfere with the target neurological phenomena). Cognition-facilitating role of body movements suggests that the restrictions might be interfering with the brain's usual routines for offloading cognitive effort, with potential effects on cognitive performance and ecological validity.[4] Thus, unless eliminating motor execution is explicitly specified as a design rationale, researchers can aim to identify data artifacts during data processing instead of imposing strict restrictions during observation (Gramann et al., 2014; Makeig et al., 2009), and even be cautious with prior designations regarding what are data artifacts (i.e., extrinsic to task-relevant cognition). More generally, researchers should distinguish mental performance (e.g., motor imagery performance) from BCI control performance as the latter depends on experiment-specific signal acquisition and processing methods. A potential pitfall of restricting body movements is improving BCI control performance at the expense of mental performance.

### 4.5.3. *Emphasis on joint performance over information transmission*

The integration focus also makes the joint performance of the brain-plus-system a more broadly applicable evaluation criterion than the accurate transmission of prior user intentions. The latter has traditionally been prevalent in BCI research as a success metric (Wolpaw et al., 2000) but comes with conceptual and practical challenges. On a conceptual level, the concurrent and interdependent nature of cognitive processes makes it difficult to sharply delineate between entities such as an information source and a channel. On a practical level, the assumption of user intentions that exist independent of and prior to externalization becomes less productive due to 1) the challenge of defining them as ground truths in uncontrolled, everyday settings and 2) the potential move toward more complex tasks where user intentions can be inherently vague (such as when a BCI is used to compose a message at once instead of dictating it word by word).

---

[4]The methodological issue we raise is analogous to the distributed cognition framework's criticism against the exclusion of external cognitive aids in study designs (Hutchins, 1995).

Focusing on joint performance avoids these problematic aspects related to information transmission criteria. Consider the following implications. For evaluation, researchers can formulate their evaluation metrics in objective, third-person terms, without alluding to prior intentions. Subjective, first-person constructs, if employed, can be defined in terms of user acceptance. Here, BCI research can take a page out of the methodology of recommender systems and apply metrics that stand for post hoc user acceptance such as "precision" in place of classification accuracy. For design, researchers can target increasing performance even when it comes at the expense of the information transmission rate. Providing task-specific constraints is one such design strategy; it limits the range of actions available to users (and thus decreases the information value of each user input) but can increase the overall performance by embedding part of the task-related information into the interface.[5] Another example is modeling mental states based on other information sources (e.g., the environment or data from other users) instead of relying exclusively on real-time brain signals.

### 4.5.4. *Expansion of the BCI research scope*
Finally, the integration perspective calls for studying BCIs in terms of their role in configuring brain activity, which implies an expansion of research scope to longer-term adaptations related to BCI use. This is particularly applicable to research that targets able-bodied users for whom relying on already existing motor skills might not lead to radically different functionality and experience than provided by current input devices. For example, many recent BCIs (e.g., Blankertz et al., 2007; Makin et al., 2020; Willett et al., 2021) exploit existing motor imagery patterns in the brain to reduce training time and improve performance. Yet alternative interfaces such as an EEG-controlled robotic third arm (Penaloza & Nishio, 2018) can extend body schema and enable new action opportunities. This counterintuitively suggests that BCIs can be valuable even if they lag in traditional performance criteria provided that they lead to qualitatively different interaction outcomes. To observe these qualitative differences, researchers can opt for open-ended tasks in user studies (e.g., by instructing text composition in place of text transcription).

The aspects of adaptation and skill-building, in return, emphasize integration qualities that are not obvious when BCIs are seen as neutral ways of accessing prior mental content. For example, a level of persistence in BCI behavior is likely required for new skills and activity patterns to develop. Consider how BCIs show different levels of persistence. Intracortical BCIs (e.g., Hochberg et al., 2012) are permanently attached while other scalp-based BCIs are temporary. Various BCIs show different levels of recording and calibration stability (Degenhart et al., 2020). Persistence in brain–interface connections has long been a focus in clinical studies (e.g., Downey et al., 2018; Perge et al., 2013), but should also be relevant for its role in supporting new forms of expression.

## 5. Reflecting on embodiment in HCI

Having identified implications for research, we find it timely to reflect on HCI's understanding of embodiment and derive implications for embodiment research at large. Below, we introduce some of the tensions within embodied interaction literature that come to the fore with the emergence of BCIs and introduce parallel efforts in cognitive science in tackling them. Of particular relevance is the necessity of body movements for cognition and attitudes toward mental phenomena.

---

[5]This example illustrates another point: Integration properties should not be interpreted as normative evaluation criteria. More bandwidth and integration within a system is not necessarily better, a point made early in systems thinking (Bunge, 1979).

## 5.1. *Embodiment without body movements*

Embodied approaches to cognition commonly state that cognition is dependent on the body. Yet various claims within embodiment emphasize different aspects, ranging from the body's role as a constraint on cognition to a mediator between the brain and the environment (R. A. Wilson & Foglia, 2017). Thus, a standing challenge for embodied cognition has been to determine this dependence. Patients with Locked-in Syndrome (LIS), who have severely restricted motor capabilities but are often cognitively intact,[6] provide a good opportunity for confronting the question (Kyselo & DiPaolo, 2015). The case of LIS patients presents a challenge for accounts of embodiment that present body movements as a prerequisite for cognition. It can, however, be accommodated by other accounts that expand the scope of embodiment to encompass simulations of motor action in the brain as well as non-motor interaction with the environment.[7] Body movements—when available —play a role in cognition, but are not a prerequisite for it (Mahon & Caramazza, 2008).

We observe that HCI understandings of embodiment contain similar tensions regarding the significance of body movements but with differences owing to HCI's constructive orientation. Some HCI interpretations of embodiment put significant emphasis on body movements (Klemmer et al., 2006), which in turn motivates utilizing a rich repertoire of physical forms and body gestures when designing interfaces. Others understand embodiment as an approach that applies to any interaction regardless of the input method (Dourish, 2004, 2013; Svanæs, 2013). The latter's emphasis is on the manipulation of the environment rather than body movements per se. That said, the brain has traditionally been reliant on the body for manipulation. As put by Dourish: "a disembodied brain could not experience the world in the same ways we do, because our experience of the world is intimately tied to the ways in which we act in it." (2004, p. 18).

The emergence of BCIs raises the question of what it means for a brain to be embodied. In this regard, BCIs present conceptual challenges to embodied interaction in the same way LIS patients pose challenges to embodied cognition, namely identifying the relevance of embodiment beyond overt body movements. Paralleling the discussions in embodied cognition, we argue for expanding the scope of embodiment to encompass internal simulations and non-motor interactions. First, HCI research can investigate how internal representations that are based on interface use are employed *even when there is no real-time interaction*, in the same way off-line cognition can involve sensorimotor simulations in the brain (M. Wilson, 2002). Second, as an alternative to a strict body–interface distinction, future research can study the functional parity of various corporal and artificial structures that come into play during cognition (Clark & Chalmers, 1998). The extent a BCI becomes body-like depends on its ability to provide comparable highly-coupled, persistent, and individualized connections, and is thus a matter of integration.

## 5.2. *Different dimensions of embodiment and mental phenomena*

We noted early on in the paper that embodiment in cognitive science and other disciplines can better be described as a collection of different theoretical and methodological commitments that individual studies only selectively commit to. For example, neurally grounded, connectionist models of cognition explain human behavior by observing neural structures (e.g., Smolensky, 1988), but can keep environmental interactions out of their scope. In contrast, researchers working in the tradition of distributed cognition study the interdependence between mental processing and interactions with the environment (e.g., Kirsh & Maglio, 1994). One way to make sense of these diverse approaches is to view them within a continuum of different "dimensions of embodiment" that vary in physical and time scale (Rohrer, 2007). Common among different dimensions is a commitment to empirical grounding, but individual studies' units of observation can range from neural structures in the brain to large organizational systems with multiple individuals. Determining the dimension of embodiment and the unit of observation is a practical question that depends on one's research purpose.

---

[6]Although there is evidence that motor imagination, which is frequently used in BCIs, is impaired in LIS patients (Conson et al., 2008).

[7]Prior work (Kyselo & DiPaolo, 2015) in particular underlined the ability of the enactivist tradition (Varela et al., 2017) in defining embodiment without delimiting its scope to the standard definitions of the body.

We can now reflect on how different lines of research related to embodied interaction position themselves along this scale. Looking back at some influential work that shaped embodied interaction discourse in HCI (Marshall & Hornecker, 2013), we observe that the neural level of embodiment attracted relatively less attention. For example, phenomenology-informed work studies phenomena such as transparency during tool use. Yet this is done mainly from the first-person viewpoint instead of grounding the experience at the neural level. Some other frameworks deliberately avoid mental phenomena by focusing on what is externally observable. Situated action approach builds its analysis on the sequential relationships between agents' observable actions and avoids making inferences about mental states (Suchman, 1993). Its research methods of choice, conversation/interaction analysis, observe the minute details of speech, gaze, and other actions that come into play during interaction, but the neural basis of social intelligibility (investigated by e.g., Gallese & Lakoff, 2005; Niedenthal, 2007; Tomasello et al., 2005) is excluded from the scope of observation. Distributed cognition leads to a similar outcome by expanding cognition's unit of analysis to the environment. In Hutchins' words: "with the new unit of analysis, many of the representations can be observed directly, so in some respects, this may be a much easier task than trying to determine the processes internal to the individual that account for the individual's behavior." (1995, p. 266). The decision to avoid mental phenomena can be justified through explanatory parsimony; brain activity is often inaccessible to researchers and model-heavy approaches run the risk of making spurious assumptions. As sensing technologies make brain activity more accessible, however, we expect the mental models in HCI to gain further empirical grounding. Thus, the current line of HCI research that uses brain-sensing for non-interactive, research purposes (for a review see Putze et al., 2022) can be informed by insights and methods from embodied cognition to explain phenomena such as tool extension (Bassolino et al., 2010; Bergström et al., 2019) or virtual presence in neural terms.

## 6. Conclusion

Early in the paper, we noted that it is possible to frame the promise of a technology in multiple ways depending on the particulars of a research vision. We then pointed to the limitations of framing the potential utility of BCIs in terms of bypassing the body and directly accessing mental states. The many interdependencies between the wider body and cognition in the brain instead demonstrate a level of integration that rebuts the conception of the body as a mere communication channel for prior mental content.

Acknowledging this integration has multiple implications for BCI research. Most obviously, future research should utilize the structure of the wider body and motor skills as a resource for guiding the evaluation and design of BCIs. Some research directions we proposed are providing body-grounded explanations for BCI performance, designing BCIs in ways that preserve body mappings, and even incorporating various sensorimotor variables and tangibles into BCI use. These research directions gain particular prominence as BCI applications are increasingly being designed for able-bodied users, a shift that also requires researchers to more clearly justify their rationale for deploying BCIs. A more abstract implication for research is the application of systemic thinking that defines the sensorimotor integration within the body to the brain–computer interaction itself. This calls for studying the role of interfaces in structuring the cognitive processes within the brain. Instead of treating various BCIs as different outlets for the same mental content, we would expect them to facilitate distinct capabilities and outcomes. An alternative, embodiment-compatible BCI research vision can thus be defined as *constructing new cognitive integrations*, which emphasizes future BCIs' role in shaping brain activity as opposed to being a neutral means for accessing mental content.

In parallel, the understanding of embodiment within HCI needs to evolve in order to be relevant for BCI research. Following previous work in cognitive science, we argued for going beyond embodied interaction's current "externalist" focus and encompassing internal simulations as well as BCI-mediated interactions. As such, embodiment is better viewed as a set of commitments that apply to the study of any input modality rather than directly making a case for using body movements over BCIs. The expanded focus also opens new research directions for the broader embodiment research within HCI. The scope of user studies can expand beyond real-time interaction to examine how past interactive experiences shape off-line cognition in the brain. As opposed to departing from prior body–interface divisions, researchers can study the functional parity of different entities at the neural level. Observations related to embodiment can be explained in more detail by identifying their neural correlates.

The particular research directions we proposed are by no means exhaustive or final. Future HCI research will likely be informed by ongoing developments in cognitive and neurosciences while at the same time testing the validity of knowledge generated in these fields for interactive applications. HCI can uniquely contribute to an embodied focus in BCI through its constructive skills, broad methodological toolbox, and ability to situate BCIs among a wider array of input devices and interaction styles. Looking forward, we consider the emergence of BCIs a grand practical and intellectual challenge for HCI and hope that our paper facilitates further reflection in the field.

## Acknowledgments

This work was supported by the European Commission under the grants 826266 (CO-ADAPT) and 824128 (VIRTUALTIMES). We would like to thank Stephen Fairclough, Tapio Takala, Antti Salovaara, and the anonymous reviewers for their feedback on the draft versions of this paper.

## Funding

The work was supported by the Horizon 2020 Framework Programme [824128,826266].

## Disclosure statement

No potential conflict of interest was reported by the authors.

## Notes on contributors

*Barış Serim* received his PhD from Aalto University in 2020; his research interests are multimodal interaction, adaptive interfaces and models of interaction in HCI.

*Michiel Spapé* received his PhD in Psychology from Leiden University in 2009 and is now docent in cognitive neuroscience at Helsinki University, focusing on emotion, perception/action, and EEG.

*Giulio Jacucci* is a professor in the Department of Computer Science at the University of Helsinki; his research interests include multimodal interaction for search and information discovery, physiological computing, and ubiquitous computing for behavioral change.

## ORCID

Barış Serim  http://orcid.org/0000-0003-1553-9642
Michiel Spapé  http://orcid.org/0000-0002-3126-1380
Giulio Jacucci  http://orcid.org/0000-0002-9185-7928